# How many software engineering professionals hold this certificate?

**Fedor Dzerzhinskiy (Dzerjinski) [1]**
**(fdzer@acm.org)**

***Abstract:*** *Estimates of quantity of the certificates issued during 10 years of existence of the professionals certification program in the area of software engineering implemented by one of the leading professional associations are presented. The estimates have been obtained by way of processing certificate records openly accessible at the certification program web-site. Comparison of these estimates and the known facts about evolution of the certification program indicates that as of the present day this evolution has not led to a large scale issuance of these certificates. But the same estimates, possibly, indicate that the meaning of these certificates differs from what is usually highlighted, and their real value is much greater. Also these estimates can be viewed, besides everything else, as reflecting an outcome of a decade long experimental verification of the known idea about "software engineering as a mature engineering profession," and they possibly show that this idea deserves partial revision.*

### *1. Introduction*

The basic subject of consideration in this article is one of existing professional certification programs in the field of software and systems engineering. At present, there are quite a lot such programs. Some of them are provided by various software vendors, consulting companies, training centers, universities, frequently in connection with the training courses in some specialized subdomains of the discipline - such as software project management, requirements, testing, etc.

Here we consider a certification program of other kind. It was implemented 10 years ago by one of the leading professional associations in the area of computing. The program is intended to cover the discipline as a whole, not just a subset of its specialized subareas, is "neutral" regarding both the commercial products of various vendors, and the existing methodological and theoretical "schools" in the field of software and systems engineering.

During all time of existence of the certification program, despite the periodically asked questions, the exact numbers of the certificates earned, breakdown of certificate issuance by country or year, how many of them are still valid (while their holders need to renew them, get recertified each three years), etc. were never explicitly reported.

This article presents the estimates of the corresponding numbers. The estimates have been obtained by way of processing the records openly accessible via certificate-holders-search page of the certification program web-site.

---

[1] OJSC "Promsvyazbank", Moscow, Russia. All stated in this material reflects the author's personal viewpoint, not necessarily identical to that of his employer.
This version 4b of the article reflects the certificate issuance statistics as of January 2012..

In Roundtable materials, Nov. 2012 (see http://www.labrate.ru/20121120/stenogramma.htm), current versions of this article in English (translation by the author) and in Russian are located at http://www.arxiv.org/abs/1211.4347 .

The material of the article is arranged as follows. After some general comments (section 2), a minimum necessary remarks are presented about the certification program being discussed (section 3), and its reform in 2008 (section 4). Further, section 5, briefly explains the ways of obtaining the source data on the basis of which the estimates presented were calculated. In section 6 the obtained estimates of the number of certificate holders are presented and commented on. Sections 7 - 10 deal with the interpretation of the obtained estimates. Section 11 contains conclusions. The main conclusions are about the value of the discussed certification and reasons why greater attention to it and active participation in it of professionals might be justified.

All stated in this paper reflects the author's personal viewpoint, not necessarily identical to that of any organization.

### *2. General remarks*

The data on the basis of which the estimates presented in this article were calculated, are for sure not absolutely exact. In particular, the processing of this data revealed presence of incomplete entries (their number is reflected below) and the discrepancies in the separate fields. From time to time some corrections of records occur on the site. Also, the information about the certification program presented below might be oversimplified, incomplete and can contain occasional inaccuracies. This information, accumulated from a variety of open sources, didn't undergo sufficient independent verification.

Since the presented information cannot be regarded as completely authentic, the exact names of the discussed certificates, and the organization that issues them are not being mentioned here. The two kinds of certificates being discussed are mentioned here as "Professional" and "Associate", the professional association that issues them - as "the Certification Provider", its team administering the certification - as "the Certification Team".

For some of the readers the information presented here can be sufficient to unambiguously identify the certification program being discussed. If such reader would wish, she/he even could independently obtain for the corresponding certification program similar estimates, and check the accuracy of results presented here.

If similar or more detailed statistics are explicitly published by the Certification Provider, such data will supercede the data presented here.

This article, any references to it, or to the data it includes shall not be used in ways which can cause harm to any certification program (similar or not similar to that discussed here), its developers, certification team, certificate holders, certification candidates, or any professional association. To the extent permitted by the applicable copyright law, this requirement shall be interpreted as part of the legal terms and conditions of lawful use of this article.

### *3. Certification program*

3.1. Originally, the certification program offered a single type of the certificate mentioned here as "Professional". A candidate had to demonstrate the fulfillment of the eligibility requirements, to pay a fee (a few hundred dollars), and to pass an examination (test) for the mastery of the core Body of Knowledge of software engineering.

3.2. Eligibility requirements originally included the following:

(1) A bachelor university degree or equivalent, at a minimum.

(2) Proof of professionalism including:

a) adherence to a Code of ethics and professional conduct [1],

b) Membership in the Certification Provider organization, or in the other recognized professional association; or the written recommendation from two members of the Certification Provider organization; or registration as a professional engineer (the requirements described in this sub-item were canceled briefly after being instituted).

(3) Experience - about 6 years of work in 6 of the specified 11 Knowledge Areas (KAs) of software engineering (see below). A detailed experience report had to be presented in the established form, including the information about institution, assignments, dates and number of hours.

3.3. The first KA in the list of 11 KAs of software engineering was: "Professionalism and engineering economics". (Subareas - engineering economics; ethics; professional practice, including legal issues; standards. Later, in the Graduate Software Engineering Curriculum GSwE2009 [2] corresponding issues of ethics and "professionalism" also constituted the first KA "Ethics and Professional Conduct", and "Engineering Economics" was included as an element in the KA "Software Engineering Management"; see [2], appendices C.1 and C.3.)

Names of the other Knowledge Areas correspond to 10 KAs of software engineering according to SWEBOK Guide 2004 [3] (they were also inherited in GSwE2009 [2]): Software Requirements; Software Design; Software Construction; Software Testing; Software Maintenance; Software Configuration Management; Software Engineering Management; Software Engineering Process; Software Engineering Tools and Methods; Software Quality.

3.4. An examination was a 3.5 hour long computer-based test in which the candidate had to answer 180 questions covering all of the 11 Knowledge Areas listed above. A few dozen sample questions with answers for self-preparation for the test were accessible on a website of the certification program and in additional materials accessible by candidates. A ratio of correct answers necessary for passing the test was unique for each variant of the test and has been never disclosed. (One can guess, it is not less than 70 %.)

3.5. The certificate "Professional" has the limited validity period - 3 years. By the end of this period the certificate holder has to get "recertified" by either reporting the prescribed amount of "units" earned during this period by certain categories of activities aimed at continuing education and professional development, or passing the certification exam again. In both cases a certificate holder has also to pay, correspondingly, recertification fee or examination fee.

### *4. Reform of the certification program*

4.1. The certification program existed without significant changes compared to the described above since 2001 till the middle of 2007. (Although the program has been officially approved by its Provider in January, 2002, the first certificate holders have taken the test in 2001 in the course of its "beta-testing".) In 2007 composition of the Certification Team has changed, and the program has undergone an essential reform.

One of the reasons for the certification program reform was probably a discrepancy between expected and actual quantity of certificates issued and, as a consequence, between expected and actual financial results of the certification program that became apparent. Possibly, by analogy to the most successful experience in other areas some persons influencing a policy of the Provider expected that there will be many thousands of valid certificates, and the certification program will become a good source of income for the Provider. Actually, on the average, less than 100 certificates per year are being issued, and holders of many of them do not renew them by the end of a 3 year validity period (see below, in section 6).

Consideration of only the "economics" aspects of the initial version of certification program (without paying attention to specificity of such categories, as "profession" and "professionalism" in the field of software engineering) reveals the presence of an evident factor, which limits the "growth of sales" of the discussed certificates. This factor is rather strict requirements the certificate holders must satisfy. A natural step to increase the sales is this: Ease the requirements, and thus expand a circle of buyers.

4.2. In line with this logic, in 2007-2008 in addition to the certificate "Professional", intended for experienced professionals, the "facilitated" variant of certification being named here as "Associate" has been developed and approved, which is intended for entry-level specialists - for bachelors without an experience, and even for the undergraduate students capable to pass an exam on the mastery of Body of Knowledge of software engineering (though at a less "deep" level, than it it is required from a "Professional"). The certificate "Associate" does not expire after 3 years.

*NOTE:* It is necessary to notice, that a target of the "Associate" certification was not just individual candidates, but primarily universities in Asian countries with rapidly-growing demand for software professionals, such as India and China.[2] The partnership with the Provider allows universities to propose the "Associate" examination as an exit test for their computer science and software engineering undergraduate students. This allows both a successful candidate and a university to demonstrate a level of competence according to internationally recognized professional standards. Being a mechanism, which allows universities in different countries to "calibrate" in the above way their educational programs in software engineering, the discussed certification can be of separate great value regardless of immediate financial outcome and absolute number of issued certificates.

4.3. In the course of the reform the boundaries of Body of Knowledge underlying the certification have been essentially reconsidered with orientation for an audience of undergraduate students and entry-level specialists, instead of experienced professionals. In particular, three new Knowledge Areas have been added, which correspond in SWEBOK 2004 [3] not to Knowledge Areas of software engineering, but to "Related Disciplines": Computing Foundations, Mathematical Foundations, Engineering Foundations. The first Knowledge Area dealing with Professionalism and Engineering Economics, was divided into two parts placed closer to the end of the list, after Software Quality.

4.4. The above change, intended to suit the needs of an "Associate" certification was also accompanied by similar change of requirements for the "Professional" certificate holders in the direction of easing these requirements. These changes also promoted adoption of more "fuzzy", less "focused" view of a concept "software engineering" in the certification program.

The easing of eligibility requirements for a "Professional" has consisted in replacement of previous definition of required experience (a specified number of thousands of hours - about 6 years - in 6 of 11 specified Areas) by a smaller number of hours (approximately on a quarter) in the "software engineering/development" (without references to specific KAs). And in some cases the experience requirement is reduced (in particular, if the candidate has an advanced degree in software engineering). The education requirements were reduced too, a candidate for the "Professional" certification, who holds an "Associate" certificate is not required to hold a bachelor degree.

### *5. The source data*

The lists of all certificate holders and of the "new" ones who have obtained it in the past half-year have been accessible and were being updated periodically on the website of the certification program since 2002 till the middle of 2007 (according to an accompanying note, the lists were not necessarily complete). Then the updates of these lists have ceased to occur, and later the lists were removed out of the open access.

---

[2] Many thanks to S. who pointed this out to the author.

After some time, instead of lists a form for the "certificate holder search" was introduced at the certification program website. The form permits to specify a few first letters of a surname. The result of the search is a table with the list of certificate holders, whose surnames begin with the specified Latin letters. Columns of the table specify person's name, country, state, city, kind of the certificate ("Professional" or "Associate"), date of the certification, current status of the certificate holder - "Active" (i. e., certificate is valid), or "Inactive".

The shortest search pattern is two letters (there are the surnames consisting of two letters). If one searches one-by-one for all pairs of Latin letters (the apostrophe too has to be regarded as a letter, there are the surnames beginning with " O ' ") it is possible to consecutively look through and (if one has prepared, for example, some small scripts in PERL) to process the entire list of certificate holders as of current date.

This is that very way the estimates presented below were obtained. The estimates are based on the data as of January, 2012. Pairs of letters for which the result of a search was nonempty: *abdefghjlmnprstuwyz, baehiloruy, cahilorsu, daehioruyz, eadilnpstvy, faeiloru, gaehiloruw, haeiouy, icgmnsv, jaeiou, kaehilnoruw, laeilouy, macehioruy, naeiou, o'cgklmnrsuw, paehiloru, qa, raehiouy, sacehiklmoptuwyz, taehioru, ulmnpr, vaeiloy, waehioru, xiu, yaeiou, zaehiuvy*. (This is not meant to be read aloud.)

*NOTE:* There are very few cases (about 1%) when, probably, the same person has earned two certificates "Professional", of which the second has been issued already after the expiry of the first. For simplicity, here we consider each such person as two certificate holders.

### *6. Estimates of number of certificate holders*

The obtained estimates of number of certificate holders are presented in the tables below. (A row of a table which contains the "????" corresponds to the incomplete entries in the source data.)

6.1. Table 1 and Table 2 contain the summary data regarding all the period of existence of the certification program. The following discussion of the data reflected in the tables, is purely qualitative, and is not meant to be regarded as indisputable.

As can be seen, for both kinds of the certificate there were "less fruitful" and "more "fruitful" years (when a "usual" level was surpassed in some times). At least the following among the "fruitful" years coincide with certain events due to which more certificates have been issued during these years.

All holders of the certificates "Professional" with dates of issuance in 2001, many of those who obtained these certificates at the end of 2009, and also the majority of holders of the certificates "Associate" with dates in 2007 and 2008 are the participants of "beta-testing" of the corresponding certification examinations. (The next beta-testing of the updated version of examination "Associate" with participation of 200 volunteers was planned to be held at the end of summer - beginning of autumn 2011. There was a growth of number of "Associates" in 2011, but not as large as in 2008.)

There were other years, in each of which more than 100 certificates "Professional" have been issued - 2003, 2004. At least one of them, 2003 is a year when active promotion of the discussed certification program was carried out at large conference for software developers in the USA.

Except the "fruitful" years when the growth was possibly caused by the above-mentioned events, the tables show that the "usual" (in the absence of such events) number of the certificates "Professional" issued yearly during the last 6 years has been stable or slightly decreased, remaining in the range of about 54-68 per year. The reform of the certification program has not affected this noticeably - neither positively, nor negatively. Also, as can be seen, more than half of the issued certificates are already not valid - their 3-year term has expired, and their holders have not renewed them by means of getting recertified.

Regarding the certificates "Associate", Table 2 demonstrates that their number per half a year over the period since July, 2008 till December, 2011 following a decline at the beginning up to the recent time grew with acceleration: 19, 9, 10, 15, 22, 43, 45. But only in the first half of 2011 their number has become greater than the number of certificates "Professional" issued during the same half a year, and remains so far relatively small.

6.2. Table 3 shows the distribution of the issued certificates over the countries. Taking into account that in the USA much more certificates, than anywhere else, have been received, it is also interesting to estimate "concentration" of the certificates in separate USA states. The corresponding data are given in Table 4.

For the same reason, in Table 1 the number of certificates "Associate" issued in the USA has been singled out in a separate column. As can be seen, only in 2011 their number became a little more than a half of all issued. Till this year, as opposed to a situation with certificates "Professional", demand for certificates "Associate" in the USA was significantly less than a total over the other countries (among them it is greatest in India and China).

**Table 1. Estimates of the certificate issuance per year.**
**(As of January, 2012: "All" - issued that "Year"; "Active" - of them, on January, 2012; "% Active " - relative to "All" for the specified "Year".)**

| **Year** | **"Professional"** | | | **"Associate"** | | |
|---|---|---|---|---|---|---|
| | **All** | **Active** | **% Active** | **All** | **USA** | **% USA** |
| 2001 | **192** | **49** | 25% | | | |
| 2002 | **16** | **3** | 18% | | | |
| 2003 | **148** | **42** | 28% | | | |
| 2004 | **134** | **39** | 29% | | | |
| 2005 | **75** | **16** | 21% | | | |
| 2006 | **55** | **14** | 25% | | | |
| 2007 | **60** | **20** | 33% | **5** | 4 | 80% |
| 2008 | **56** | **26** | 46% | **158** | 73 | 46% |
| 2009 | **153** | **153** | 100% | **19** | 5 | 26% |
| 2010 | **54** | **54** | 100% | **37** | 11 | 29% |
| 2011 | **68** | **68** | 100% | **87** | 53 | 60% |
| ???? | 6 | 6 | 100% | 14 | 2 | 14% |
| **Total** | **1017** | **490** | **48%** | **320** | **148** | **46%** |

**Table 2. Estimates of the certificate issuance per month.**
**(As of January, 2012. Totals till 2008 - for a year, since 2008 - for half a year)**

| Year - month | "Pro-fess." | "As-soc." |
|---|---|---|
| 2001-04 | 8 | |
| 2001-05 | 107 | |
| 2001-06 | 77 | |
| **2001** | **192** | |
| 2002-01 | 1 | |
| 2002-05 | 1 | |
| 2002-06 | 14 | |
| **2002** | **16** | |
| 2003-04 | 5 | |
| 2003-05 | 52 | |
| 2003-06 | 91 | |
| **2003** | **148** | |
| 2004-04 | 29 | |
| 2004-05 | 17 | |
| 2004-06 | 40 | |
| 2004-09 | 2 | |
| 2004-10 | 10 | |
| 2004-11 | 36 | |
| **2004** | **134** | |
| 2005-01 | 6 | |
| 2005-04 | 17 | |
| 2005-05 | 2 | |
| 2005-06 | 15 | |
| 2005-10 | 5 | |
| 2005-11 | 30 | |
| **2005** | **75** | |
| 2006-04 | 1 | |
| 2006-05 | 4 | |
| 2006-06 | 24 | |
| 2006-09 | 1 | |
| 2006-10 | 6 | |
| 2006-11 | 19 | |
| **2006** | **55** | |

| Year - month | "Pro-fess." | "As-soc." |
|---|---|---|
| 2007-04 | 2 | |
| 2007-05 | 2 | |
| 2007-06 | 24 | |
| 2007-09 | 2 | |
| 2007-10 | 6 | |
| 2007-11 | 14 | |
| 2007-12 | 10 | 5 |
| **2007** | **60** | **5** |
| | | |
| 2008-01 | 1 | 136 |
| 2008-02 | 1 | 3 |
| 2008-04 | 2 | 0 |
| 2008-05 | 2 | 0 |
| 2008-06 | 4 | 0 |
| **2008 I** | **10** | **139** |
| 2008-07 | 12 | 1 |
| 2008-09 | 2 | 8 |
| 2008-11 | 13 | 2 |
| 2008-12 | 19 | 8 |
| **2008 II** | **46** | **19** |
| 2009-01 | 2 | 2 |
| 2009-02 | 2 | 2 |
| 2009-03 | 1 | 0 |
| 2009-04 | 2 | 0 |
| 2009-05 | 1 | 1 |
| 2009-06 | 7 | 4 |
| **2009 I** | **15** | **9** |
| 2009-07 | 18 | 2 |
| 2009-08 | 1 | 0 |
| 2009-10 | 2 | 1 |
| 2009-11 | 10 | 2 |
| 2009-12 | 107 | 5 |
| **2009 II** | **138** | **10** |

| Year - month | "Pro-fess." | "As-soc." |
|---|---|---|
| 2010-01 | 1 | 1 |
| 2010-02 | 1 | 0 |
| 2010-03 | 12 | 0 |
| 2010-04 | 0 | 8 |
| 2010-05 | 2 | 3 |
| 2010-06 | 5 | 3 |
| **2010 I** | **21** | **15** |
| 2010-07 | 1 | 3 |
| 2010-08 | 3 | 0 |
| 2010-09 | 4 | 2 |
| 2010-10 | 0 | 1 |
| 2010-11 | 4 | 3 |
| 2010-12 | 21 | 13 |
| **2010 II** | **33** | **22** |
| 2011-01 | 1 | 0 |
| 2011-02 | 1 | 0 |
| 2011-03 | 3 | 1 |
| 2011-04 | 1 | 1 |
| 2011-05 | 4 | 4 |
| 2011-06 | 13 | 36 |
| **2011 I** | **23** | **42** |
| 2011-07 | 0 | 2 |
| 2011-08 | 2 | 1 |
| 2011-09 | 2 | 24 |
| 2011-10 | 7 | 1 |
| 2011-11 | 7 | 1 |
| 2011-12 | 27 | 16 |
| **2011 II** | **45** | **45** |
| ???? | 6 | 15 |
| **Total** | **1017** | **320** |

**Table 3. Breakdown of the certificate issuance by countries.**
**(As of January, 2012)**

| "Professional" | | | "Associate" | |
|---|---|---|---|---|
| All | Active | | | Country |
| **753** | **347** | **46%** | **148** | **USA** |
| 53 | 33 | 62% | 20 | Canada |
| 34 | 26 | 76% | 52 | India |
| 26 | 7 | 26% | 0 | Korea |
| 25 | 9 | 36% | 40 | China |
| 13 | 7 | 53% | 4 | Germany |
| 11 | 1 | 9% | 0 | Poland |
| 8 | 2 | 25% | 4 | Brazil |
| 7 | 4 | 57% | 5 | United Kingdom |
| 6 | 2 | 33% | 8 | Australia |
| 6 | 3 | 50% | 2 | Japan |
| 6 | 5 | 83% | 2 | Spain |
| 6 | 3 | 50% | 0 | Taiwan |
| 5 | 0 | 0% | 1 | Singapore |
| 4 | 0 | 0% | 0 | Argentina |
| 4 | 3 | 75% | 5 | Egypt |
| 3 | 3 | 100% | 2 | Mexico |
| 3 | 2 | 66% | 0 | New Zealand |
| 3 | 2 | 66% | 4 | South Africa |
| 3 | 3 | 100% | 0 | Sweden |
| 2 | 1 | 50% | 0 | Chile |
| 2 | 0 | 0% | 0 | Denmark |
| 2 | 2 | 100% | 0 | Hong Kong |
| 2 | 2 | 100% | 1 | Italy |
| 2 | 2 | 100% | 0 | Malaysia |
| 2 | 1 | 50% | 1 | Netherlands |
| **2** | **1** | **50%** | **4** | **Russia** |
| 2 | 2 | 100% | 1 | Switzerland |
| 1 | 1 | 100% | 0 | Belgium |
| 1 | 1 | 100% | 0 | Colombia |
| 1 | 1 | 100% | 0 | Cyprus |
| 1 | 0 | 0% | 0 | Dominican Republic |
| 1 | 1 | 100% | 0 | France |
| 1 | 1 | 100% | 0 | Ghana |
| 1 | 1 | 100% | 1 | Greece |
| 1 | 0 | 0% | 0 | Hungary |
| 1 | 1 | 100% | 0 | Ireland |
| 1 | 0 | 0% | 0 | Israel |
| 1 | 1 | 100% | 0 | Kuwait |
| 1 | 1 | 100% | 0 | Latvia |
| 1 | 1 | 100% | 0 | Norway |
| 1 | 1 | 100% | 0 | Peru |
| 1 | 1 | 100% | 0 | Saudi Arabia |
| 1 | 0 | 0% | 0 | Slovakia |
| 1 | 1 | 100% | 1 | United Arab Emirates |
| 1 | 1 | 100% | 0 | Yugoslavia |
| 0 | 0 | | 1 | Austria |
| 0 | 0 | | 1 | Croatia |
| 0 | 0 | | 1 | Guatemala |
| 0 | 0 | | 2 | Lebanon |
| 0 | 0 | | 1 | Morocco |
| 0 | 0 | | 1 | Nigeria |
| 0 | 0 | | 1 | Pakistan |
| 0 | 0 | | 2 | Uganda |
| | | | | |
| 1 | 1 | 100% | 2 | ???? |

## Table 4. Estimates of the certificate issuance in the US states.
## (As of January, 2012)

| "Professional" | | | "Associate" | |
|---|---|---|---|---|
| All | Active | | | State |
| **753** | **347** | **46%** | **148** | **Total in the USA** |
| 90 | 34 | 37% | 17 | CALIFORNIA |
| 54 | 29 | 53% | 17 | TEXAS |
| 53 | 25 | 47% | 5 | VIRGINIA |
| 47 | 16 | 34% | 3 | WASHINGTON |
| 40 | 21 | 52% | 9 | FLORIDA |
| 38 | 21 | 55% | 2 | MASSACHUSETTS |
| 36 | 18 | 50% | 7 | MARYLAND |
| 29 | 15 | 51% | 5 | NEW JERSEY |
| 29 | 14 | 48% | 8 | PENNSYLVANIA |
| 28 | 11 | 39% | 3 | MICHIGAN |
| 27 | 12 | 44% | 0 | ILLINOIS |
| 24 | 15 | 62% | 11 | NEW YORK |
| 24 | 6 | 25% | 5 | ALABAMA |
| 21 | 9 | 42% | 3 | OHIO |
| 19 | 5 | 26% | 4 | COLORADO |
| 18 | 4 | 22% | 14 | INDIANA |
| 17 | 6 | 35% | 1 | MINNESOTA |
| 16 | 11 | 68% | 1 | GEORGIA |
| 15 | 6 | 40% | 3 | SOUTH CAROLINA |
| 13 | 6 | 46% | 4 | ARIZONA |
| 12 | 9 | 75% | 3 | NORTH CAROLINA |
| 11 | 6 | 54% | 1 | NEW MEXICO |
| 11 | 1 | 9% | 2 | UTAH |

| "Professional" | | | "Associate" | |
|---|---|---|---|---|
| All | Active | | | State |
| 7 | 5 | 71% | 0 | NEW HAMPSHIRE |
| 7 | 5 | 71% | 9 | WISCONSIN |
| 7 | 4 | 57% | 2 | OREGON |
| 6 | 4 | 66% | 0 | CONNECTICUT |
| 6 | 3 | 50% | 1 | IOWA |
| 5 | 5 | 100% | 1 | Washington, DC |
| 5 | 2 | 40% | 0 | KANSAS |
| 5 | 2 | 40% | 0 | MISSOURI |
| 5 | 2 | 40% | 2 | TENNESSEE |
| 4 | 3 | 75% | 1 | IDAHO |
| 4 | 1 | 25% | 0 | ARKANSAS |
| 3 | 2 | 66% | 0 | NEBRASKA |
| 3 | 0 | 0% | 1 | NEVADA |
| 2 | 2 | 100% | 0 | VERMONT |
| 2 | 1 | 50% | 1 | MAINE |
| 2 | 1 | 50% | 0 | MISSISSIPPI |
| 2 | 1 | 50% | 0 | RHODE ISLAND |
| 1 | 1 | 100% | 0 | ALASKA |
| 1 | 1 | 100% | 0 | MONTANA |
| 1 | 0 | 0% | 0 | DELAWARE |
| 1 | 0 | 0% | 0 | HAWAII |
| 0 | 0 | | 1 | OKLAHOMA |
| 0 | 0 | | 1 | WEST VIRGINIA |
| 2 | 2 | 100% | 0 | ???? |

### *7. Optimistic interpretation of the estimates obtained*

As frequently happens, one and the same estimates can provide a basis for quite different interpretations. We shall begin with their optimistic interpretation.

The estimates provide a ground to hope that a steady, accelerated growth of the number of certificates "Associate" will continue in the future.

Also there are factors, other than these estimates, which provide a ground for an optimism. In particular:

- The fact that financial outcomes of the Certification Program are determined not by the quantity of issued certificates, but by the quantity of examinations, as well as preparation courses and materials paid for, and this quantity is many times greater.

- The fact that large number of organizations, including a number of the largest commercial, as well as government structures recommend that their employees should hold such a certificate. An extensive list of such organizations is accessible at the Web-site of Certification Program, and it steadily grows.

- Also grows the number of universities and training centers in different countries, which have joined the Certification Program as Registered Education Providers, which are entitled to provide preparation courses for the examination.

The above mentioned and other factors allow to argue that there exists a favorable prospect, including the financial one, for this certification program, we have just to wait for some years more.

Using this argumentation it is possible to disregard as economically insignificant the following facts about the certificates "Professional":

- The maximum of number of valid certificates "Professional" has remained in the remote past. As on January 31, 2005 - the expiry date of the first issued certificates, there were 496 valid certificates; at the end of 2011 - 490.

- Less than a half of the issued certificates are valid - 48%.

- The total number of such certificates issued for more than 10 years, only at the end of 2011 slightly exceeded 1000.

### *8. Cautious doubt*

The growth of number of the certificates "Associate" and other factors mentioned above really provide a certain ground for optimistic point of view. But there are reasons to cautiously doubt that this point of view is well-grounded in all respects.

Although there is a rapid growth of number of certificates "Associate", for many years their total number still remains very small. At the same time, both for these certificates, and for certificates "Professional" there are "fruitful" years when, in a comparable period, much more certificates are issued, than "usually". This is caused by certain circumstances of irregular character, and without taking them into account the consideration of dynamics of the certification program hardly will be adequate.

Examples of such circumstances are promotion of testing at conferences and especially, "beta-testing" of the examination. In such cases the participants of certification usually have additional moral motivation (assistance to development of the profession, etc.), and also have some discounts and privileges. The discount for participants of "beta-testing" in 2001 of the very first version of examination was especially significant. At that time the certification fee was a fraction of "usual" fee. It is quite in line with the record number of certificates "Professional" that year, which has never been repeated.

The other kind of circumstances influencing the rates of growth of number of certificates is a participation of universities and training centers in the certification program. The participation level, whether an initial enthusiasm will be replaced by disappointment and indifference at a later point depends on many individual factors.

In view of such reasons, let's consider, alongside with the above "optimistic" interpretation of the discussed estimates, their other interpretation **as an evidence** that the certification program **produced the results that differed substantially from the expected ones**.

### *9. Valuable experimental result*

This interpretation cannot be regarded as "pessimistic". The science knows well that the more unexpected result of an experiment is obtained, the more valuable this result can be. This certification program apart from anything is a serious experiment that has been conducted already for a decade. It can be helpful to realize that in the course of this experiment a really unexpected result (that is, potentially valuable) has been obtained. It is not a ground for pessimism, but an occasion to attempt finding reasonable explanations for the experimental result obtained.

To discuss the certification program from this point of view it can be useful to divert our attention for a while away from its aspects related to its "financial outcome", and ask ourselves questions of another sort: What number of these certificates "had" to be issued? In view of financial aspects of the certification program (which, certainly, has demanded appreciable expenses), undoubtedly expectations existed of much greater certificate issuance than it turned out in the reality. On the basis of what explicit or implicit assumptions was this expected, and which of these initial "hypotheses" have not proved to be true? Something was wrong in the initial model of the phenomenon, if the model generated the expectations which didn't come true. What was wrong in this model?

The initial conceptual model underlying the discussed certification program is one of interpretations of idea about "software engineering as a mature engineering profession" [4]. Possible ways of updating this model (and its later interpretations), including the ways of its updating based on the "experimental results", similar to ones mentioned above, is too serious issue to discuss it here systematically.

But one point nevertheless seems to be worth being noted.

### *10. One of the initial hypotheses: A middle between a specialist and a generalist?*

The model of "software engineering as a profession" attempted to single out characteristic attributes (components) of the recognized "mature" professions, such as professions of physician, lawyer or civil engineer. Software creation activities have significant consequences for the society. Recognition of this has led to the idea that the society should regulate these activities through the formation of a mature "engineering" profession of software engineering with similar attributes.

The components of an "engineering profession", as considered in [4], are briefly summarized, in particular, in the Introduction of SWEBOK Guide 2004 ([3], page 1-1). Their list includes, as important components, voluntary certification or mandatory licensing for confirming fitness of the professional to practice in the field of software engineering. This includes demonstration of a mastery of the Body of Knowledge, agreed upon by a broad consensus within the professional community, regarding what a software engineering professional should know.

SWEBOK Guide is an attempt to define this Body of Knowledge. Initially, an underlying Body of Knowledge of the discussed certification program, as was mentioned above, was similar to an extended version of the SWEBOK (extensions are the "professional practices" and "engineering economics", treated similarly to GSwE2009 [2]). Following the update in 2009, the certification program has been declared to be based upon SWEBOK (without a reference to specific version), all extensions in it remained in place. (Inclusion of a reference to SWEBOK was caused, probably, by an aspiration to formally conform to the ISO standard which has appeared by that time [5].)

The domain of the considered Body of Knowledge consists of specialized narrow subdomains. In most cases a specialist in some of these subdomains to work in then productively and safely, not necessarily should know all other subdomains at the same thorough level. Even more, the "congenital" individual qualities of the person, which are favorable for his/her most effective work in some subdomains, can be incompatible with qualities, which are optimal in other subdomains (see [7]).

The general methodological problem, which is probably hard to avoid while implementing any certification program similar to discussed here, consists in a difficulty of balancing between requirements to the competences of a specialist (depth of knowledge in separate particular subdomains) and of a generalist (coverage of all essential subdomains and the discipline as a whole).

S.Optner in his book [6] wrote the following regarding specialists and generalists:

"*According to the philosopher, Ralph Burton Perry, a specialist is a person who knows more and more about less and less, until he knows almost everything about nothing. Conversely he defines a generalist as a person who knows less and less about more and more, until he knows practically nothing about everything.*"

The considered certificate "Professional" is intended for experienced professionals, and the variant of the solution "nothing about everything" in this case would not be of practical value.

Developers of the certification program have attempted to solve a non-trivial problem of finding "the golden mean" - "about everything", but also "a lot of". (In this connection there was even coined a special term - "a midlevel software engineer". This is, roughly speaking, the one who, being a generalist, contrives to not become "a pure manager", remains a competent technical specialist, and in many knowledge areas. This term for a long period was mentioned in promotional materials, its meaning remained quite unclear in that context. Finally the "mysterious" concept "midlevel" has been replaced with "midcareer", and then it was further replaced with simply "experienced".)

The definition of certification requirements (not only examination) based on the this approach has been developed by a process of broad consensus within the collective of the recognized experts, highly experienced representatives of professional community. But in view of marked above non-trivial nature of the task, any its chosen solution could not be anything else but a hypothesis subject to check by practice. The above quantitative data, probably are showing, that this hypothesis has not proved to be true.

Namely, it is quite possible, that they demonstrate **an absence of the bases to hope the certificate "Professional" (or similar to it) will be obtained by a significant number of the practicing professionals in the field of software engineering**. It is unattainable and **because of this** it is not necessary - for most of them (for experts in various areas, including the software project management), for the market, for the society, for the state.

But the same data, probably, bring to light the existence of a special, and rather rare sort of professionals in different countries. Characteristic of them is the ability not only to demonstrate conformity to requirements of the certificate "Professional", but also to understand, why this certificate is necessary for them, and to understand this as much definitely, as to invest many hours and days of time and an appreciable enough sum of money in its obtaining. (Actually, the proved comprehension of it is one more, "latent", but possibly one of the major criteria of belonging to this sort of professionals.)

The meaning and value of belonging of the professional to this group of "elite" can be most clear to colleagues who already belong to it (including "inactive" - those whose certificates have expired), or to those having prerequisites to enter it (including, in the long-term future or even if they are not going to enter into it). But also for more general public, though the certificate is rather mysterious, it possesses quite significant "weight", caused mostly by extremely small number ("rarity") of its holders, by the presence among them of persons who are rather eminent within the professional community, and also, not least, by the well established brand names of the professional association and the certificate.

Maybe the most unexpected result, shown by the above quantitative data, concerns the certificate "Associate". This result is that, if to judge only by the quantity, at present this certificate is not less, and maybe even more valuable (in its own "age category"), than the certificate "Professional", and exactly for the same reasons: as the entrance ticket in the same "elite" group.

Substantial research of the characteristic traits of this sort of professionals, of possible roles of such professionals in various organizations and processes (and not only of creation, but also of the use of software), would be very interesting, but is beyond the boundaries this consideration.

One other point worth to be noted here, is that the distribution of number of certificate holders over the countries (see Table 3) can serve in itself as a ground for considerations.

### *11. Instead of the conclusion: About the value of mistakes*

Many years ago the author has had the luck to attend a lecture of C.A.R.Hoare at a scientific seminar in the Lomonosov Moscow State University (it was when E.W.Dijkstra and C.A.R.Hoare have visited Moscow in 1976 [8]). After the lecture one of the attendees has asked the lecturer to comment on an opinion that the path to understanding of "true nature" of one of the central categories of an area considered in the lecture (as far as I remember it was a concept of concurrency) will be "long, continuous sequence of ridiculous mistakes". Professor Hoare has replied, that he agreed with that opinion, and "would have liked to make most of these mistakes personally".

The problems, pointed out in this article concerning the incongruity between the expectations and the actual results cannot serve a ground for doubts regarding the value of certification programs under consideration. On the contrary, in view of these results the value of such programs can be essentially higher than what is usually attributed to them.

And even these very problems can constitute an unexpected and consequently potentially valuable experimental result of practical verification of adequacy of the currently popular views on the nature of such categories, as "profession" and "professionalism" in the field of software engineering.

Quite probably, many of efforts on reforming the considered certification programs will seem in the future as the mistakes on a path to understanding of "true nature" of such categories. In view of the Professor Hoare's above-mentioned idea, these mistakes are valuable, and those who make them, deserve a respect and gratitude.

For those readers who, in connection with the above, would be interested in examples of more detailed information about certification programs similar to the one discussed here, the author would like to recommend references [9, 10].

***Acknowledgments***

The author is grateful to all persons who have reviewed preliminary versions of this paper, for their attention, valuable feedback, help, and support.

arXiv:1211.4347v4 [cs.SE] 2012 (edited 2017)

**How many software engineering professionals hold this certificate?**
**Fedor Dzerzhinskiy (Dzerjinski)**

***Abstract:*** *Estimates of quantity of the certificates issued during 10 years of existence of the professionals certification program in the area of software engineering implemented by one of the leading professional associations are presented. The estimates have been obtained by way of processing certificate records openly accessible at the certification program web-site. Comparison of these estimates and the known facts about evolution of the certification program indicates that as of the present day this evolution has not led to a large scale issuance of these certificates. But the same estimates, possibly, indicate that the meaning of these certificates differs from what is usually highlighted, and their real value is much greater. Also these estimates can be viewed, besides everything else, as reflecting an outcome of a decade long experimental verification of the known idea about "software engineering as a mature engineering profession," and they possibly show that this idea deserves partial revision.*

# Сколько профессионалов программной инженерии имеют этот сертификат?

**Федор Янович Дзержинский [1]**

**(fdzer@acm.org)**

***Аннотация:*** *Представлены оценки количества сертификатов, выданных на протяжении 10 лет существования программы сертификации профессионалов в области программной инженерии, реализованной одной из ведущих профессиональных ассоциаций. Приводимые оценки получены путем обработки записей о выданных сертификатах, общедоступных на веб-сайте программы сертификации. Сопоставление с этими оценками происходившей эволюции рассматриваемой программы сертификации показывает, что на сегодня эта эволюция не привела к крупномасштабному выпуску этих сертификатов. Но эти же оценки, возможно, указывают на иное, чем обычно подразумевается, смысловое значение рассматриваемых сертификатов и их намного большую реальную ценность. Также эти оценки можно рассматривать, помимо всего прочего, как отражение результата десятилетней экспериментальной проверки известной идеи о "программной инженерии как зрелой инженерной профессии", и возможно они указывают на необходимость частичного пересмотра этой идеи.*

---

[1] ОАО "Промсвязьбанк", г. Москва. Все изложенное в данном материале отражает личную точку зрения автора, не обязательно совпадающую с точкой зрения работодателя.
Данная редакция 4b статьи отражает статистику выпуска сертификатов по состоянию на **январь 2012 г**.

В материалах Круглого стола, ноябрь 2012 г. (http://www.labrate.ru/20121120/stenogramma.htm, сетевой ресурс), текущие версии статьи на английском (перевод автора) и на русском языке - http://www.arxiv.org/abs/1211.4347

### *1. Введение*

Основным предметом обсуждения в данной статье является одна из существующих программ сертификации профессионалов в области программной и системной инженерии. На сегодня известно довольно много подобных программ. Часть из них реализуется различными поставщиками ПО, консалтинговыми, учебными центрами, университетами, часто в связи с учебными курсами, касающимися тех или иных специализированных разделов дисциплины - таких как управление программными проектами, управление требованиями к программному обеспечению (ПО), тестирование ПО и т. п.

Здесь рассматривается программа сертификации иного рода. Она реализована 10 лет назад одной из ведущих профессиональных ассоциаций в области вычислительной техники, ориентирована на охват всей области знаний, а не отдельных ее специализированных подобластей, "нейтральна" по отношению как к коммерческим продуктам различных поставщиков, так и к существующим методологическим и теоретическим "школам" в области программной и системной инженерии.

На протяжении всего времени существования рассматриваемой программы сертификации, несмотря на периодически задаваемые вопросы, ни разу не были объявлены какие-либо претендующие на точность фактические данные о количестве выданных сертификатов, о том, сколько их выдано в какой стране, в разные годы, сколько из них действуют (с учетом необходимости продления - "ресертификации" каждые три года) и т. п.

В данной статье представлены оценки соответствующих данных, полученные путем несложной обработки записей, общедоступных посредством страницы поиска обладателей сертификатов на Веб-сайте программы сертификации.

Материал статьи организован следующим образом. После ряда пояснений общего характера (раздел 2), приведен необходимый минимум сведений о рассматриваемой программе сертификации (раздел 3) и ее реформе в 2008 г. (раздел 4). Далее, в разделе 5, кратко пояснены способы получения исходных данных, на основе которых вычислены приводимые в статье оценки. В разделе 6 представлены и пояснены полученные оценки количества обладателей сертификатов. В разделах 7 - 10 изложен ряд соображений, касающихся интерпретации рассматриваемых оценок. Заключение содержит выводы, основной из которых - о ценности рассматриваемой сертификации и о целесообразности большего внимания к ней и активного участия в ней профессионалов.

Все изложенное в данной статье отражает личную точку зрения автора, не обязательно совпадающую с точкой зрения какой-либо организации.

### *2. Пояснения общего характера*

Исходные данные, на основе которых вычислены приводимые в этой статье оценки, заведомо не являются абсолютно точными. На это указывают, в частности, обнаружившееся при обработке этих данных наличие неполных записей (их число ниже указано), неточностей в отдельных полях, время от времени происходящие уточнения записей на сайте. Кроме того, приводимые ниже сведения о программе сертификации вынужденно упрощены, неполны и могут содержать случайные неточности. Эта информация, накопленная из разнообразных открытых источников, не проходила достаточной независимой выверки.

Поскольку приводимые данные не могут рассматриваться как полностью достоверные, здесь не упоминаются точные названия ни рассматриваемых сертификатов, ни выдающей их организации. Два вида рассматриваемых сертификатов условно упоминаются здесь как "Профессионал" и "Ассоциат", выдающая их профессиональная ассоциация - как

"Провайдер сертификации", ее структура, администрирующая сертификацию - как "Команда сертификации".

Для многих из читателей излагаемых здесь сведений может оказаться достаточно для того, чтобы однозначно идентифицировать программу сертификации, о которой здесь идет речь. При желании, такой читатель мог бы самостоятельно получить для соответствующей программы сертификации аналогичные оценки и проверить результаты, приводимые здесь.

Если аналогичная или более подробная статистика будет опубликована Провайдером сертификации, она сделает устаревшими приводимые здесь данные в их части, расходящейся с этой статистикой.

Эта статья, ссылки на нее и приводимые в ней данные не должны использоваться способами, могущими нанести вред какой-либо программе сертификации (похожей или не похожей на рассматриваемую здесь), ее разработчикам, команде сертификации, обладателям сертификатов, кандидатам на получение сертификатов, любой профессиональной ассоциации. В той степени, в которой возможно в соответствии с применимым законодательством об авторском праве, это требование должно рассматриваться как существенное условие законного использования настоящего литературного произведения.

### *3. Программа сертификации*

3.1. Первоначально, в рамках программы сертификации выдавался лишь один вид сертификата, здесь условно именуемый "Профессионал". Кандидат на его получение должен был подтвердить свое соответствие ряду "квалификационных" (eligibility) требований, заплатить определенную сумму (несколько сотен долларов) и пройти экзамен (тест) на владение "совокупностью знаний" (Body of Knowledge) программной инженерии.

3.2. Квалификационные требования первоначально включали следующее:

(1) Наличие университетского диплома не ниже бакалавра или его эквивалента.

(2) Подтверждение профессионализма, включавшее:

(2а) принятие Кодекса этики и профессионального поведения [1],

(2б) членство в организации - Провайдере сертификации, либо в признанной ею иной профессиональной организации, либо письменная рекомендация от двух членов организации - Провайдера сертификации, либо наличие статуса зарегистрированного профессионального инженера (требования, описанные в этом подпункте, были вскоре отменены).

(3) Опыт - около 6 лет работы в 6 из 11 перечисленных "Областей знания" (Knowledge Area, KA) программной инженерии (см. ниже). Об опыте требовалось представить детальные сведения по установленной форме, с указанием мест работы, ее характера и количества часов.

3.3. Перечень 11 Областей знания программной инженерии начинался с Области "Профессионализм и инженерная экономика". (Подобласти - инженерная экономика; этика; профессиональная практика, включая правовые вопросы; стандарты. В изданных позднее рекомендациях по магистерским программам программной инженерии GSwE2009 [2] соответствующие вопросы этики и "профессионализма" также составляют первую Область знаний "Этика и профессиональное поведение", а "инженерная экономика" включена как элемент в Область знаний "Менеджмент инжиниринга ПО"; подробнее о них см. [2], приложения C.1 и C.3).

Названия остальных Областей соответствуют 10 Областям знаний программной инженерии согласно руководству SWEBOK Guide 2004 [3] (они повторены и в GSwE2009 [2]): Требования к ПО; Проектирование ПО; Конструирование ПО; Тестирование ПО;

Сопровождение ПО; Конфигурационное управление ПО; Менеджмент инжиниринга ПО; Процесс инжиниринга ПО; Инструменты и методы инжиниринга ПО; Качество ПО.

3.4. Экзамен представлял собой компьютерный тест, в рамках которого за 3,5 часа кандидат должен был ответить на 180 вопросов, охватывавших все перечисленные выше Области. Примерно 100 примеров вопросов с ответами для самоподготовки к тесту были доступны на сайте программы сертификации и в дополнительных материалах для кандидатов. Количество вопросов, на которые нужно ответить для того, чтобы успешно пройти тест, - свое для каждого варианта теста и не раскрывалось. (Можно предположить, что оно не меньше 70%.)

3.5. Сертификат "Профессионал" имеет ограниченный срок действия - 3 года. К концу этого срока обладатель сертификата должен пройти "ресертификацию" либо отчитавшись о том, что он за прошедшее время набрал заданное количество "зачетных единиц", присваиваемых за определенные категории деятельности по непрерывному обучению и профессиональному развитию, либо повторно сдав сертификационный экзамен. В обоих случаях обладатель сертификата должен также оплатить, соответственно, ресертификацию или экзамен.

***4. Реформа программы сертификации***

4.1. Программа сертификации просуществовала без значительных изменений по сравнению с тем, как она охарактеризована выше, с 2001 г. до середины 2007 г. (Хотя программа была официально одобрена ее Провайдером в январе 2002 г., первые обладатели сертификатов прошли тест в 2001 г. в рамках его "бета-тестирования".) В 2007 г. сменился состав Команды сертификации, и программа претерпела существенные изменения.

Одной из их причин явилось, вероятно, ставшее очевидным расхождение между ожидавшимся и фактическим количеством выданных сертификатов и, как следствие, между ожидавшимися и фактическими финансовыми результатами программы сертификации. Возможно, по аналогии с наиболее успешным опытом в других областях некоторые из лиц, влияющих на политику Провайдера ожидали, что число действующих сертификатов достигнет многих тысяч, и программа сертификации станет хорошим источником дохода для Провайдера. Фактически же оказалось, что выдается в среднем меньше 100 сертификатов в год, и что многие обладатели сертификатов не продлевают их по истечении 3-летнего срока действия (см. ниже, в разделе 6).

Если рассматривать первоначальную программу сертификации в чисто "экономическом" ракурсе (не обращая внимания на специфику таких категорий, как "профессия" и "профессионализм" в области программной инженерии), то бросается в глаза наличие фактора, ограничивающего "рост продаж" рассматриваемых сертификатов. Этот фактор - довольно строгие требования к кандидатам на получение сертификата. Естественный шаг к увеличению продаж - снизить требования, тем самым расширив круг покупателей.

4.2. В соответствии с этой логикой, в 2007-2008 гг. в дополнение к предназначенному для опытных профессионалов сертификату "Профессионал" был разработан и утвержден "облегченный" вариант сертификации, условно именуемый здесь "Ассоциат", и предназначенный для начинающих специалистов - бакалавров без опыта работы, а также для студентов, еще не ставших бакалаврами, но способных сдать экзамен на владение совокупностью знаний программной инженерии (хотя и на менее "глубоком" уровне, чем требуется от "Профессионала"). Сертификат "Ассоциат" не имеет срока действия.

*ПРИМЕЧАНИЕ.* Необходимо отметить, что целевой аудиторией сертификации "Ассоциат" были не только индивидуальные кандидаты на получение сертификата, но и прежде всего университеты в азиатских странах с быстро растущей потребностью в профессионалах ПО, таких как Индия и Китай.[2] Партнерство с Провайдером позволяет университетам предлагать экзамен "Ассоциат" в качестве заключительного экзамена для своих студентов бакалавриата в области информатики и программной инженерии. Это помогает и успешному кандидату, и университету демонстрировать уровень компетентности согласно международно признанным профессиональным стандартам. Будучи механизмом, позволяющим университетам в различных странах указанным образом "откалибровать" свои программы обучения в области программной инженерии, рассматриваемая сертификация может иметь самостоятельную большую ценность, независимо от непосредственных финансовых результатов и абсолютного количества выданных сертификатов.

4.3. В процессе реформы границы лежащей в основе сертификации совокупности знаний программной инженерии были существенно пересмотрены с ориентацией на аудиторию студентов и начинающих специалистов, а не опытных профессионалов. В частности, в нее были добавлены три новых Области, соответствующие отнесенным в SWEBOK 2004 [3] не к Областям знаний программной инженерии, а к "связанным с ней" дисциплинам: Основы вычислений, Математические основы, Инженерные основы. Первая Область, касающаяся Профессионализма и Инженерной экономики, была разбита на две части, помещенные ближе к концу списка, после Области Качество ПО.

4.4. Изменение в расчете на "Ассоциата" используемого в программе сертификации взгляда на совокупность знаний программной инженерии сопровождалось и изменением требований к обладателям сертификата "Профессионал" в сторону их ослабления. Эти изменения также повлекли некоторое "размывание", снижение "сфокусированности" лежащего в основе программы сертификации понятия "программная инженерия".

Ослабление квалификационных требований к "Профессионалу" выразилось в замене прежнего определения требуемого опыта (столько-то тысяч часов - около 6 лет - в 6 из 11 указанных Областей) требованием меньшего количества часов (примерно на четверть) в области "инжиниринга/разработки ПО" (без конкретизации по Областям). Причем в ряде случаев требование к опыту сокращено (в частности, при наличии степени магистра или выше в области программной инженерии). Также снижено требование к образованию, кандидат на получение сертификата "Профессионал", имеющий сертификат "Ассоциат", может не быть бакалавром.

### *5. Исходные данные*

С 2002 до середины 2007 г. на Веб-сайте программы сертификации были доступны и периодически обновлялись списки всех обладателей сертификата, а также "новых" - получивших его в истекшем полугодии (было оговорено, что списки могут быть неполными). Затем эти списки перестали обновлять, позднее доступ к ним вообще закрыли.

Спустя какое-то время, вместо списков на сайте программы сертификации появилась экранная форма для поиска обладателей сертификатов по первым буквам фамилии. Результат поиска - таблица со списком обладателей сертификатов, чьи фамилии начинаются с указанных букв латиницы. В столбцах таблицы указаны страна, штат, город, вид сертификата ("Профессионал" или "Ассоциат"), дата сертификации, текущий статус сертификата - "Активный" (т. е., действующий), либо "Неактивный".

Минимальный образец для поиска - две буквы (существуют фамилии, состоящие из двух букв). Если перебрать все пары букв латиницы (апостроф тоже нужно считать буквой, есть фамилии, начинающиеся с "O'"), то можно последовательно просмотреть и (подготовив, например, небольшие скрипты на PERL) обработать весь список обладателей сертификатов по состоянию на текущую дату.

[2] Большое спасибо S. за то, что он обратил внимание автора на данный момент.

Таким способом и получены приводимые ниже оценки. Оценки основаны на данных по состоянию на Январь 2012 г. Пары букв, результаты поиска которых были непустыми: *abdefghjlmnprstuwyz, baehiloruy, cahilorsu, daehioruyz, eadilnpstvy, faeiloru, gaehiloruw, haeiouy, icgmnsv, jaeiou, kaehilnoruw, laeilouy, macehioruy, naeiou, o'cgklmnrsuw, paehiloru, qa, raehiouy, sacehiklmoptuwyz, taehioru, ulmnpr, vaeiloy, waehioru, xiu, yaeiou, zaehiuvy*. (Это не предназначено для чтения вслух.)

*ПРИМЕЧАНИЕ.* Есть очень немного случаев (около 1%), когда, вероятно, один и тот же человек получил два сертификата "Профессионал", второй из которых выдан уже после истечения действия первого. Для простоты, здесь мы рассматриваем каждого такого человека как двух обладателей сертификата.

### *6. Оценки количества обладателей сертификатов*

Полученные оценки количества обладателей сертификатов представлены в приводимых ниже таблицах. (Строки таблиц, в которых указано "????" соответствует неполным записям исходных данных.)

6.1. Таблица 1 и Таблица 2 содержат сводные данные за весь период существования программы сертификации. Приводимые ниже пояснения к отраженным в таблицах данным имеют сугубо качественный характер и не претендуют на бесспорность.

Как видно из таблиц, для обоих видов сертификата были менее "урожайные" годы и более "урожайные" (когда "обычный" уровень превышался в несколько раз). По крайней мере следующие из "урожайных" годов можно соотнести с определенными событиями, способствовавшими тому, что в эти годы было выдано больше сертификатов.

Обладатели сертификатов "Профессионал", датированных 2001 г., а также многие из получивших такие сертификаты в конце 2009 г. и большинство получивших сертификаты "Ассоциат" в 2007 и 2008 г. - это участники "бета-тестирования" соответствующих сертификационных экзаменов. (Очередное бета-тестирование обновленной версии экзамена "Ассоциат" с участием 200 добровольцев было намечено на конец лета - начало осени 2011 г. Рост числа "Ассоциатов" в 2011 г. имел место, но менее значительный, чем в 2008 г.)

Были другие годы, в каждом из которых было получено больше 100 сертификатов "Профессионал" - 2003, 2004. По крайней мере один из них, 2003 - это год, когда проводилось активное продвижение рассматриваемой программы сертификации в рамках большой конференции в США для разработчиков ПО.

Из таблиц видно, что за исключением "урожайных" лет, когда рост был возможно вызван указанными событиями, "обычное" (при отсутствии подобных событий) количество выдаваемых за год сертификатов "Профессионал" в последние 6 лет стабилизировалось или незначительно снижалось, оставаясь в пределах около 54-68 за год. Реформа программы сертификации на него заметно не повлияла - ни в лучшую, ни в худшую сторону. Также видно, что более половины выданных сертификатов уже не действуют - утратили силу по истечении 3-летнего срока, и их обладатели не продлили их путем "ресертификации".

Что касается сертификатов "Ассоциат", то из Таблицы 2 видно, что за период с июля 2008 г. по декабрь 2011 г. их количество за полугодие, после небольшого спада вначале, вплоть до самого недавнего времени увеличивалось с ускорением: 19, 9, 10, 15, 22, 42, 45. Но лишь в первом полугодии 2011 г. оно стало больше количества полученных за тот же период сертификатов "Профессионал", и остается пока относительно небольшим.

6.2. В Таблице 3 представлено распределение выданных сертификатов по странам. Поскольку в США их выдано намного больше, чем где-либо еще, представлялось интересным оценить также "концентрацию" сертификатов в отдельных штатах США. Соответствующие данные приведены в Таблице 4.

По той же причине в Таблице 1 отдельно выделено количество сертификатов "Ассоциат", выданных в США. Видно, что только в 2011 г. их стало чуть больше половины от общего количества выданных. До этого года, в противоположность ситуации с сертификатами "Профессионал", спрос на сертификаты "Ассоциат" в США был существенно меньше, чем суммарно в остальных странах (среди них он наибольший в Индии и Китае).

**Таблица 1. Оценки количества обладателей сертификатов по годам. (По состоянию на январь 2012 г.: "Все" - выданные в указанный "Год"; "Активные" - из них, на январь 2012 г.; "% Активных" - по отношению ко "Всем" за указанный "Год".)**

| Год | "Профессионал" | | | "Ассоциат" | | |
|---|---|---|---|---|---|---|
| | Все | Актив-ные | % Ак-тивных | Все | США | % США |
| 2001 | **192** | **49** | 25% | | | |
| 2002 | **16** | **3** | 18% | | | |
| 2003 | **148** | **42** | 28% | | | |
| 2004 | **134** | **39** | 29% | | | |
| 2005 | **75** | **16** | 21% | | | |
| 2006 | **55** | **14** | 25% | | | |
| 2007 | **60** | **20** | 33% | **5** | 4 | 80% |
| 2008 | **56** | **26** | 46% | **158** | 73 | 46% |
| 2009 | **153** | **153** | 100% | **19** | 5 | 26% |
| 2010 | **54** | **54** | 100% | **37** | 11 | 29% |
| 2011 | **68** | **68** | 100% | **87** | 53 | 60% |
| ???? | 6 | 6 | 100% | 14 | 2 | 14% |
| **Всего** | **1017** | **490** | **48%** | **320** | **148** | **46%** |

**Таблица 2. Оценки количества выданных сертификатов по месяцам.
(По состоянию на январь 2012 г. Итоги до 2008 г. - за год, с 2008 г. - за полугодие)**

| Год - месяц | "Про-фесс." | "Ас-соц." |
|---|---|---|
| 2001-04 | 8 | |
| 2001-05 | 107 | |
| 2001-06 | 77 | |
| **2001** | **192** | |
| 2002-01 | 1 | |
| 2002-05 | 1 | |
| 2002-06 | 14 | |
| **2002** | **16** | |
| 2003-04 | 5 | |
| 2003-05 | 52 | |
| 2003-06 | 91 | |
| **2003** | **148** | |
| 2004-04 | 29 | |
| 2004-05 | 17 | |
| 2004-06 | 40 | |
| 2004-09 | 2 | |
| 2004-10 | 10 | |
| 2004-11 | 36 | |
| **2004** | **134** | |
| 2005-01 | 6 | |
| 2005-04 | 17 | |
| 2005-05 | 2 | |
| 2005-06 | 15 | |
| 2005-10 | 5 | |
| 2005-11 | 30 | |
| **2005** | **75** | |
| 2006-04 | 1 | |
| 2006-05 | 4 | |
| 2006-06 | 24 | |
| 2006-09 | 1 | |
| 2006-10 | 6 | |
| 2006-11 | 19 | |
| **2006** | **55** | |

| Год - месяц | "Про-фесс." | "Ас-соц." |
|---|---|---|
| 2007-04 | 2 | |
| 2007-05 | 2 | |
| 2007-06 | 24 | |
| 2007-09 | 2 | |
| 2007-10 | 6 | |
| 2007-11 | 14 | |
| 2007-12 | 10 | 5 |
| **2007** | **60** | **5** |
| | | |
| 2008-01 | 1 | 136 |
| 2008-02 | 1 | 3 |
| 2008-04 | 2 | 0 |
| 2008-05 | 2 | 0 |
| 2008-06 | 4 | 0 |
| **2008 I** | **10** | **139** |
| 2008-07 | 12 | 1 |
| 2008-09 | 2 | 8 |
| 2008-11 | 13 | 2 |
| 2008-12 | 19 | 8 |
| **2008 II** | **46** | **19** |
| 2009-01 | 2 | 2 |
| 2009-02 | 2 | 2 |
| 2009-03 | 1 | 0 |
| 2009-04 | 2 | 0 |
| 2009-05 | 1 | 1 |
| 2009-06 | 7 | 4 |
| **2009 I** | **15** | **9** |
| 2009-07 | 18 | 2 |
| 2009-08 | 1 | 0 |
| 2009-10 | 2 | 1 |
| 2009-11 | 10 | 2 |
| 2009-12 | 107 | 5 |
| **2009 II** | **138** | **10** |

| Год - месяц | "Про-фесс." | "Ас-соц." |
|---|---|---|
| 2010-01 | 1 | 1 |
| 2010-02 | 1 | 0 |
| 2010-03 | 12 | 0 |
| 2010-04 | 0 | 8 |
| 2010-05 | 2 | 3 |
| 2010-06 | 5 | 3 |
| **2010 I** | **21** | **15** |
| 2010-07 | 1 | 3 |
| 2010-08 | 3 | 0 |
| 2010-09 | 4 | 2 |
| 2010-10 | 0 | 1 |
| 2010-11 | 4 | 3 |
| 2010-12 | 21 | 13 |
| **2010 II** | **33** | **22** |
| 2011-01 | 1 | 0 |
| 2011-02 | 1 | 0 |
| 2011-03 | 3 | 1 |
| 2011-04 | 1 | 1 |
| 2011-05 | 4 | 4 |
| 2011-06 | 13 | 36 |
| **2011 I** | **23** | **42** |
| 2011-07 | 0 | 2 |
| 2011-08 | 2 | 1 |
| 2011-09 | 2 | 24 |
| 2011-10 | 7 | 1 |
| 2011-11 | 7 | 1 |
| 2011-12 | 27 | 16 |
| **2011 II** | **45** | **45** |
| ???? | 6 | 15 |
| **Всего** | **1017** | **320** |

**Таблица 3. Оценки количества обладателей сертификатов по странам.**
**(По состоянию на январь 2012 г.)**

| "Профессионал" | | | "Ассоциат" | |
|---|---|---|---|---|
| Все | Активные | | | Страна |
| **753** | **347** | **46%** | **148** | **США** |
| 53 | 33 | 62% | 20 | Канада |
| 34 | 26 | 76% | 52 | Индия |
| 26 | 7 | 26% | 0 | Корея |
| 25 | 9 | 36% | 40 | Китай |
| 13 | 7 | 53% | 4 | Германия |
| 11 | 1 | 9% | 0 | Польша |
| 8 | 2 | 25% | 4 | Бразилия |
| 7 | 4 | 57% | 5 | Великобритания |
| 6 | 2 | 33% | 8 | Австралия |
| 6 | 3 | 50% | 2 | Япония |
| 6 | 5 | 83% | 2 | Испания |
| 6 | 3 | 50% | 0 | Тайвань |
| 5 | 0 | 0% | 1 | Сингапур |
| 4 | 0 | 0% | 0 | Аргентина |
| 4 | 3 | 75% | 5 | Египет |
| 3 | 3 | 100% | 2 | Мексика |
| 3 | 2 | 66% | 0 | Новая Зеландия |
| 3 | 2 | 66% | 4 | ЮАР |
| 3 | 3 | 100% | 0 | Швеция |
| 2 | 1 | 50% | 0 | Чили |
| 2 | 0 | 0% | 0 | Дания |
| 2 | 2 | 100% | 0 | Гонконг |
| 2 | 2 | 100% | 1 | Италия |
| 2 | 2 | 100% | 0 | Малайзия |
| 2 | 1 | 50% | 1 | Нидерланды |
| **2** | **1** | **50%** | **4** | **Россия** |
| 2 | 2 | 100% | 1 | Швейцария |

| "Профессионал" | | | "Ассоциат" | |
|---|---|---|---|---|
| Все | Активные | | | Страна |
| 1 | 1 | 100% | 0 | Бельгия |
| 1 | 1 | 100% | 0 | Колумбия |
| 1 | 1 | 100% | 0 | Кипр |
| 1 | 0 | 0% | 0 | Доминиканская Республика |
| 1 | 1 | 100% | 0 | Франция |
| 1 | 1 | 100% | 0 | Гана |
| 1 | 1 | 100% | 1 | Греция |
| 1 | 0 | 0% | 0 | Венгрия |
| 1 | 1 | 100% | 0 | Ирландия |
| 1 | 0 | 0% | 0 | Израиль |
| 1 | 1 | 100% | 0 | Кувейт |
| 1 | 1 | 100% | 0 | Латвия |
| 1 | 1 | 100% | 0 | Норвегия |
| 1 | 1 | 100% | 0 | Перу |
| 1 | 1 | 100% | 0 | Саудовская Аравия |
| 1 | 0 | 0% | 0 | Словакия |
| 1 | 1 | 100% | 1 | Объединенные Арабские Эмираты |
| 1 | 1 | 100% | 0 | Югославия |
| 0 | 0 | | 1 | Австрия |
| 0 | 0 | | 1 | Хорватия |
| 0 | 0 | | 1 | Гватемала |
| 0 | 0 | | 2 | Ливан |
| 0 | 0 | | 1 | Марокко |
| 0 | 0 | | 1 | Нигерия |
| 0 | 0 | | 1 | Пакистан |
| 0 | 0 | | 2 | Уганда |
| | | | | |
| 1 | 1 | 100% | 2 | ???? |

**Таблица 4. Оценки количества обладателей сертификатов по штатам США.**
**(По состоянию на январь 2012 г.)**

| "Профессионал" | | | "Ассоциат" | Штат |
|---|---|---|---|---|
| Все | Активные | | | |
| **753** | **347** | **46%** | **148** | Всего в США |
| 90 | 34 | 37% | 17 | Калифорния |
| 54 | 29 | 53% | 17 | Техас |
| 53 | 25 | 47% | 5 | Вирджиния |
| 47 | 16 | 34% | 3 | Вашингтон |
| 40 | 21 | 52% | 9 | Флорида |
| 38 | 21 | 55% | 2 | Массачусетс |
| 36 | 18 | 50% | 7 | Мэриленд |
| 29 | 15 | 51% | 5 | Нью-Джерси |
| 29 | 14 | 48% | 8 | Пенсильвания |
| 28 | 11 | 39% | 3 | Мичиган |
| 27 | 12 | 44% | 0 | Иллинойс |
| 24 | 15 | 62% | 11 | Нью-Йорк |
| 24 | 6 | 25% | 5 | Алабама |
| 21 | 9 | 42% | 3 | Огайо |
| 19 | 5 | 26% | 4 | Колорадо |
| 18 | 4 | 22% | 14 | Индиана |
| 17 | 6 | 35% | 1 | Миннесота |
| 16 | 11 | 68% | 1 | Джорджия |
| 15 | 6 | 40% | 3 | Южная Каролина |
| 13 | 6 | 46% | 4 | Аризона |
| 12 | 9 | 75% | 3 | Северная Каролина |
| 11 | 6 | 54% | 1 | Нью-Мексико |
| 11 | 1 | 9% | 2 | Юта |
| 7 | 5 | 71% | 0 | Нью-Хэмпшир |
| 7 | 5 | 71% | 9 | Висконсин |
| 7 | 4 | 57% | 2 | Орегон |
| 6 | 4 | 66% | 0 | Коннектикут |
| 6 | 3 | 50% | 1 | Айова |
| 5 | 5 | 100% | 1 | Вашингтон, округ Колумбия |
| 5 | 2 | 40% | 0 | Канзас |
| 5 | 2 | 40% | 0 | Миссури |
| 5 | 2 | 40% | 2 | Теннесси |
| 4 | 3 | 75% | 1 | Айдахо |
| 4 | 1 | 25% | 0 | Арканзас |
| 3 | 2 | 66% | 0 | Небраска |
| 3 | 0 | 0% | 1 | Невада |
| 2 | 2 | 100% | 0 | Вермонт |
| 2 | 1 | 50% | 1 | Мэн |
| 2 | 1 | 50% | 0 | Миссисипи |
| 2 | 1 | 50% | 0 | Род-Айленд |
| 1 | 1 | 100% | 0 | Аляска |
| 1 | 1 | 100% | 0 | Монтана |
| 1 | 0 | 0% | 0 | Делавэр |
| 1 | 0 | 0% | 0 | Гавайи |
| 0 | 0 | | 1 | Оклахома |
| 0 | 0 | | 1 | Западная Виргиния |
| 2 | 2 | 100% | 0 | ???? |

### *7. Оптимистическая трактовка полученных оценок*

Как часто бывает, в одних и те же рассматриваемых оценках можно найти основание для существенно различающихся трактовок. Начнем с их оптимистической трактовки.

Оценки дают основание надеяться, что устойчивый рост с ускорением количества сертификатов "Ассоциат" продолжится и в будущем.

Также имеются иные, чем указанные оценки, факторы, дающие повод для оптимизма. В частности:

- То, что финансовые результаты Программы сертификации определяются не количеством выданных сертификатов, а количеством оплаченных экзаменов, а также подготовительных курсов и материалов, а эти цифры многократно выше.

- То, что достаточно большое количество организаций, включая ряд крупнейших коммерческих, а также государственных структур рекомендуют при приеме на работу наличие рассматриваемого сертификата. Обширный список таких организаций имеется на сайте Программы сертификации, и он постоянно растет.

- Также растет количество университетов и учебных центров в разных странах, присоединившихся к Программе сертификации в качестве "Зарегистрированных Провайдеров Обучения" (Registered Education Providers), уполномоченных оказывать услуги по подготовке к экзамену.

Отмеченные и иные факторы позволяет обосновывать наличие благоприятной, в том числе, финансовой перспективы обсуждаемой программы сертификации, нужно лишь подождать еще несколько лет.

В этой аргументации можно игнорировать, как экономически незначащие, следующее факты, касающиеся сертификатов "Профессионал":

- Максимум числа действующих сертификатов "Профессионал" остался в далеком прошлом. На 31 января 2005 г., дату окончания срока действия первых выданных сертификатов было 496 действующих сертификатов; в конце 2011 г. - 490.

- Действуют меньше половины выданных сертификатов - 48%.

- Общее число таких сертификатов, выданных более чем за 10 лет, лишь в конце 2011 года немного превысило 1000.

### *8. Осторожное сомнение*

Рост числа сертификатов "Ассоциат" и упомянутые иные факторы действительно дают определенные основания для указанной выше "оптимистической" точки зрения. Но есть причины для осторожного сомнения в том, что она является обоснованной во всех отношениях.

При быстром росте числа сертификатов "Ассоциат", их суммарно за много лет выдано крайне мало. При этом, как для этих сертификатов, так и для сертификатов "Профессионал" характерны "урожайные" годы, когда за сравнимый период их выдается намного больше, чем "обычно". Это связано с определенными обстоятельствами нерегулярного характера, и без их учета оценка динамики рассматриваемой программы сертификации едва ли будет достоверной.

Примеры таких обстоятельств - продвижение тестирования на конференциях и "бета-тестирование" экзамена. В таких случаях участники сертификации могут иметь дополнительную моральную мотивацию (содействие развитию профессии и т. п.), а также получать те или иные скидки и льготы. Особенно значительной скидка была для участников "бета-тестирования" в 2001 г. самой первой версии экзамена. Тогда стоимость сертификации была во много раз меньше "обычной". Это вполне согласуется с не повторенным с тех пор количеством сертификатов "Профессионал" за год.

Другой вид обстоятельств, влияющих на темпы роста числа сертификатов - это подключение к программе сертификации тех или иных университетов и учебных центров. То, насколько результативным окажется подобное участие, не сменится ли первоначальный энтузиазм разочарованием и безразличием - это зависит от множества индивидуальных факторов.

С учетом подобных соображений имеет смысл, наряду с отмеченной выше "оптимистической" трактовкой рассматриваемых оценок рассмотреть другую их трактовку - как **свидетельства полного несоответствия ожидаемых и фактически полученных результатов** программы сертификации.

### *9. Ценный экспериментальный результат*

Эту трактовку было бы неправильно считать "пессимистической". В науке хорошо известно, что результат эксперимента тем ценнее, чем более неожиданным он оказывается. Рассматриваемая программа сертификации - это, помимо всего прочего, серьезный, продолжающийся уже десятилетие эксперимент. Имеет смысл отдать себе отчет в том, что ходе этого эксперимента получен по настоящему неожиданный (то есть, потенциально ценный) результат. Это - повод не для пессимизма, а для того, чтобы попытаться поискать для полученного экспериментального результата возможные разумные объяснения.

Для обсуждения программы сертификации в этом ключе может быть полезно на время отвлечься от аспектов, связанных с ее "финансовыми результатами", и задать себе вопросы иного рода: А сколько их, обладателей рассматриваемых сертификатов "должно было бы быть"? С учетом финансовых аспектов программы сертификации (она, конечно, потребовала ощутимых затрат) несомненно предполагалось, что сертификатов будет выдано намного больше, чем оказалось фактически. На основании каких явных или неявных предположений это предполагалось, и какие из этих первоначальных "гипотез" не подтвердились? Что-то было не так в исходной модели явления, если она породила не сбывшиеся ожидания. Что в ней было не так?

Исходная концептуальная модель, лежащая в основе рассматриваемой программы сертификации - это некоторая из трактовок идеи о "программной инженерии как зрелой инженерной профессии" [4]. Вопрос о возможных направлениях корректировки этой модели (и ее позднее возникших трактовок), в том числе с учетом "экспериментальных результатов", подобных упомянутым выше, - слишком серьезный для того, чтобы пытаться его здесь систематически рассмотреть.

Но один момент представляется целесообразным все же отметить.

### *10. Одна из исходных гипотез: Среднее между специалистом и дженералистом?*

В рамках модели "программной инженерии как профессии" была предпринята попытка выделить характерные признаки (компоненты) общепризнанных "зрелых" профессий, таких, как профессии медика, юриста или инженера-строителя, и поставлен вопрос о необходимости, в связи со значительными последствиями для общества деятельности по созданию ПО, регулирования этой деятельности обществом путем формирования для нее аналогичных признаков зрелой "инженерной" профессии программной инженерии.

Выделенные в работе [4] компоненты "инженерной профессии" кратко резюмированы, в частности, во введении в SWEBOK Guide 2004 ([3], стр. 1-1). Их перечень, включает в качестве одного важных компонентов добровольную сертификацию или обязательное лицензирование, подтверждающие пригодность профессионала для осуществления деятельности в области программной инженерии. При этом, в том числе, подтверждается, что он владеет определенной Совокупностью знаний (Body of Knowledge), общепризнанной на основе "широкого консенсуса" в рамках профессионального сообщества, в качестве того, что профессионал в области программной инженерии "должен знать".

SWEBOK Guide - результат попытки определить эту Совокупность знаний. Первоначальный вариант рассматриваемой программы сертификации, как было отмечено выше, основан на расширенном по сравнению со SWEBOK определении Совокупности знаний (расширения касались "профессиональных практик" и "инженерной экономики", трактуемых аналогично GSwE2009 [2]). Обновленная в 2009 г. программа сертификации считается соответствующей SWEBOK (без указания версии), все расширения в ней сохранены. (Привязка к SWEBOK обусловлена, вероятно, стремлением к формальному соответствию появившемуся к тому времени стандарту ISO [5].)

Рассматриваемая Совокупность знаний состоит из узкоспециализированных подобластей. В большинстве случаев специалист по некоторым из этих подобластей, чтобы работать в ней продуктивно и "безопасно", вовсе не обязательно должен столь же детально знать все другие подобласти. Более того, "врожденные" личные качества человека, благоприятствующие его наиболее эффективной работе в одних областях, могут быть несовместимы с качествами, оптимальными в другой подобласти (см. [7]).

Общая методологическая проблема, которой вероятно трудно избежать при реализации любой программы сертификации, подобной рассматриваемой здесь, заключается в том, что трудно найти "баланс" между требованиями к компетенции специалиста (глубина знаний в отдельных частных подобластях) и "дженералиста" (охватом всех существенных подобластей и дисциплины в целом).

С.Оптнер в книге [6] писал о специалистах и дженералистах следующее:

"*Следуя философу Ральфу Бартону Перри, специалиста можно определить как человека, который с течением времени знает все больше и больше о все-меньшем и меньшем, пока, наконец, он не станет знать почти все ни о чем. Наоборот, дженералист определяется как человек, который с течением времени знает все меньше и меньше о все большем и большем, пока, наконец, он не будет знать практически ничего обо всем.*"

Рассматриваемая программа сертификации "Профессионал" предназначена для опытных профессионалов, и вариант решения "ничего обо всем" в данном случае не имел бы практической ценности.

Разработчики программы попытались решить нетривиальную задачу - найти "золотую середину" - и "обо всем", но и "много". (В связи с этим возник даже специальный термин - "среднеуровневый программный инженер", "midlevel software engineer". Это - грубо говоря, тот, кто умудряется, будучи дженералистом, не превращаться в "чистого менеджера", остается компетентным техническим специалистом, причем во многих областях знания. Этот термин долго фигурировал в рекламных материалах, оставаясь почти для всех неясным. В итоге "таинственное" понятие "midlevel" было заменено сначала на "midcareer", т. е., что-то вроде русского "действующий", а затем просто на "опытный".)

Основанное на данном подходе определение сертификационных требований (не только экзамена) было выполнено в рамках "широкого консенсуса" коллектива признанных специалистов, многоопытных представителей профессионального сообщества. Но с учетом отмеченной выше нетривиальности поставленной задачи, любое выбранное ее решение не могло бы быть ничем, кроме гипотезы, подлежащей практической проверке. Рассмотренные выше количественные данные, вероятно, указывают, что эта гипотеза не подтвердилась.

А именно, вполне возможно, что они указывают на **отсутствие оснований надеяться на получение сертификата "Профессионал" (или подобного ему) значительным числом практикующих профессионалов в области программной инженерии**. Это недостижимо и **поэтому** не нужно - ни их основной массе (специалистам в подобластях, включая менеджмент программных проектов), ни рынку, ни обществу, ни государству.

Но эти же данные, возможно, "высвечивают" наличие особой, причем, весьма редкой разновидности профессионалов в разных странах. Ее представители характеризуются тем, что для них посильно не только продемонстрировать соответствие требованиям сертификата "Профессионал", но и осознать, зачем нужен этот сертификат, причем осознать настолько убежденно, чтобы вложить в его получение много часов и дней своего времени и достаточно ощутимую сумму денег. (Фактически, доказанное осознание этого - еще один, "скрытый", но едва ли не один из важнейших критериев принадлежности к данной разновидности профессионалов.)

Смысл и ценность принадлежности профессионала к этой группе "избранных" лучше всего понятен коллегам, к ней уже принадлежащим (в том числе и в случае прекращения действия сертификата), или имеющим предпосылки для того, чтобы в нее войти (в том числе, в отдаленном будущем или даже если они не собираются в нее входить). Но и для более широкой публики сертификат, хотя и загадочен, имеет вполне значительный "вес", обусловленный самой по себе крайней малочисленностью ("редкостью") его обладателей, наличием среди них весьма "именитых" в профессиональной среде персон, а также, не в последнюю очередь, раскрученными брендами профессиональной ассоциации и сертификата.

Возможно, самый неожиданный результат, демонстрируемый рассмотренными выше количественными данными, касается сертификата "Ассоциат". Этот результат состоит в том, что если судить только по их количеству, на данный момент этот сертификат не менее, а то и более ценен (в своей "возрастной категории"), чем сертификат "Профессионал", и в точности по тем же причинам - как входной билет в ту же самую группу "избранных".

Содержательное исследование особенностей, характерных для данной "разновидности" профессионалов, возможных ролей ее представителей в разнообразных организациях и процессах (причем не только создания, но и использования ПО), было бы весьма интересно, но выходит за рамки данного рассмотрения.

Еще один момент, который здесь представляется желательным отметить, это то, что распределение количества обладателей сертификатов по странам (см. выше Таблицу 3) может само по себе служить поводом для размышлений.

### *11. Вместо заключения: О ценности ошибок*

Много лет назад автору посчастливилось присутствовать на научном семинаре в МГУ им. М.В.Ломоносова, на котором выступил Ч.Э.Р.Хоар (когда Э.В.Дейкстра и Ч.Э.Р.Хоар ненадолго приезжали в Москву в 1976 г. [8]). После выступления один из присутствующих попросил докладчика прокомментировать мнение о том, что путь к пониманию "истинной природы" одной из центральных категорий рассматривавшейся в выступлении области (кажется, речь шла о параллелизме - concurrency) будет представлять собой "долгую, непрерывную последовательность нелепых ошибок". Профессор Хоар ответил, что с данным мнением он согласен, и что "большую часть этих ошибок хотел бы совершить лично".

Отмеченные в данной статье проблемы несовпадения ожиданий и фактических результатов отнюдь не являются поводом для сомнений в ценности программ сертификации, подобных рассмотренной. Напротив, представляется, что с учетом этих результатов ценность таких программ может быть существенно выше той, которая им обычно приписывается.

И даже сами эти проблемы могут быть неожиданным, и поэтому потенциально ценным экспериментальным результатом практической проверки адекватности распространенных на сегодня представлений о природе таких категорий, как "профессия" и "профессионализм" в области программной инженерии.

Вполне возможно, многие из усилий по реформированию рассматриваемых программ сертификации в будущем покажутся ошибками на пути к пониманию "истинной природы" подобных категорий. С учетом упомянутой выше мысли профессора Хоара, эти ошибки ценны, а те, кто их совершают, заслуживают уважения и благодарности.

Тем из читателей, у кого, в связи со всем изложенным выше, возник бы интерес к примерам более подробной информации о программах сертификации, подобных рассмотренной здесь, автор хотел бы порекомендовать ссылки [9, 10].

***Благодарность***

Автор благодарен всем, кто прокомментировал предварительную редакцию данной статьи, за их внимание, ценные замечания, помощь и поддержку.

***Литература***